\documentclass[twocolumn]{aastex631}
\usepackage{comment}  

\newcommand{\hi}{\ion{H}{1}}

\newcommand{\heii}{\ion{He}{2}}
\newcommand{\cii}{\ion{C}{2}}
\newcommand{\ciii}{\ion{C}{3}}
\newcommand{\civ}{\ion{C}{4}}

\newcommand{\nv}{\ion{N}{5}}

\newcommand{\siii}{\ion{Si}{2}}
\newcommand{\siiii}{\ion{Si}{3}}
\newcommand{\siiv}{\ion{Si}{4}}

\newcommand{\mgii}{\ion{Mg}{2}}

\newcommand{\ovi}{\ion{O}{6}}

\newcommand{\suiii}{\ion{S}{3}}
\newcommand{\oi}{\ion{O}{1}}
\newcommand{\oii}{\ion{O}{2}}
\newcommand{\oiii}{\ion{O}{3}}
\newcommand{\oiv}{\ion{O}{4}}

\begin{document}

\title{Impacts of the Metagalactic Ultraviolet Background on Circumgalactic Medium Absorption Systems}

\author[0000-0003-3886-8326]{Elias Taira}
\affiliation{Department of Physics \& Astronomy, 567 Wilson Road, Michigan State University, East Lansing, MI 48824}

\author[0000-0001-5158-1966]{Claire Kopenhafer}
\affiliation{Institute for Cyber-Enabled Research, 567 Wilson Road, Michigan State University, East Lansing, MI 48824}
\affiliation{Department of Physics \& Astronomy, 567 Wilson Road, Michigan State University, East Lansing, MI 48824}

\author[0000-0002-2786-0348]{Brian W.\ O'Shea}
\affiliation{Department of Computational Mathematics, Science, \& Engineering, Michigan State University, 428 S. Shaw Lane, East Lansing, MI 48824}
\affiliation{Department of Physics \& Astronomy, 567 Wilson Road, Michigan State University, East Lansing, MI 48824}
\affiliation{Facility for Rare Isotope Beams, Michigan State University, 640 S. Shaw Lane, East Lansing, MI 48824}
\affiliation{Institute for Cyber-Enabled Research, 567 Wilson Road, Michigan State University, East Lansing, MI 48824}

\author{Alexis Manning}
\affiliation{Department of Physics \& Astronomy, 514 University Blvd, University of Alabama, Tuscaloosa, AL 35401}

\author{Evelyn Fuhrman}
\affiliation{Department of Physics \& Astronomy, 567 Wilson Road, Michigan State University, East Lansing, MI 48824}

\author[0000-0003-1455-8788]{Molly S.\ Peeples}
\affiliation{Space Telescope Science Institute, 3700 San Martin Dr., Baltimore, MD 21218}
\affiliation{Center for Astrophysical Sciences, William H.\ Miller III Department of Physics \& Astronomy, Johns Hopkins University, 3400 N.\ Charles Street, Baltimore, MD 21218}

\author[0000-0002-7982-412X]{Jason Tumlinson}
\affiliation{Space Telescope Science Institute, 3700 San Martin Dr., Baltimore, MD 21218}
\affiliation{Center for Astrophysical Sciences, William H.\ Miller III Department of Physics \& Astronomy, Johns Hopkins University, 3400 N.\ Charles Street, Baltimore, MD 21218}

\author[0000-0002-6804-630X]{Britton D.\ Smith}
\affiliation{Institute for Astronomy, University of Edinburgh, Royal Observatory, EH9 3HJ, UK}



\begin{abstract}

Among the many different pieces of physics that go into simulations of the circumgalactic medium (CGM), the metagalactic ultraviolet background (UVB) plays a significant role in determining the ionization state of different metal species. However, the UVB is uncertain with multiple models having been developed by various research groups over the past several decades. In this work, we examine how different UVB models influence the ionic column densities of CGM absorbers. We use these UVB models to infer ion number densities in the FOGGIE galaxy simulations at $z=2.5$ and use the Synthetic Absorption Line Surveyor Application (SALSA) package to identify absorbers. Absorbers are then matched across UVB models based on their line of sight position so that their column densities can be compared. From our analysis, we find that changing the UVB model produces significant changes in ionization, specifically at lower gas densities and higher temperatures where photoionization dominates over collisional ionization. We also find that the scatter of column density differences between models tends to increase with increasing ionization energy, with the exception of \hi, which has the highest scatter of all species we examined.

\end{abstract}



\section{Introduction} \label{sec:intro}



The circumgalactic medium (CGM) is the diffuse, multiphase medium that surrounds a galaxy. It is typically observed via quasar absorption spectra due to its overall low surface brightness. Such observations have been gathered through a number of surveys, including COS-Halos \citep{tumlinson2013}, COS-Burst \citep{heckman2017}, the COS CGM Compendium \citep{lehner2018}, KODIAQ and KODIAQ Z \citep{lehner2014, lehner2016, lehner2022}, Red Dead Redemption \citep{berg2019}, CASBaH \citep{burchett2018,prochaska2019}, CUBS \citep{chen2020}, CGM2 \citep{wilde2021}, and KBSS-InCLOSE \citep{nuñez2024}. These surveys have resulted in absorption estimates for a wide range of metal ions that can then be used to extract information about the physical properties of the CGM such as its multiple gas densities, temperatures, and metallicities. 

To better understand the history of the CGM and its evolution over time, large suites of simulations have been developed that study CGM absorption structure. Their work includes cosmological simulations such as EAGLE \citep{wijers2020}, GIBLE \citep{ramesh2024}, Illustris-TNG50 \citep{defelippis2021}, FOGGIE \citep{peeples2019}, and \citet{hummels2019} as well as some idealized simulations like \citet{fielding2017} and \citet{butsky2022}. It is vital that these experiments study metal ion distributions to effectively compare the results of these simulations with observation.

Conceptually, there are two broad frameworks for connecting absorption spectra with physical gas properties: forward and reverse modeling.
Taking the physical properties of the CGM as the underlying ``ground truth,'' reverse modeling must be applied to observations in order to ``back out'' the ground-truth gas state from the observed spectra. The absorption spectrum represents a reduced set of information about the system that generated it and such a situation is also known as an ``inverse problem'' \citep[see, e.g.,][]{yaman2013survey}. Conversely, efforts to make synthetic observations of simulations are engaging in forward modeling. Synthetic observations can go as far as generating mock spectra targeting a specific instrument, such as the Hubble Space Telescope Cosmic Origins Spectrograph, complete with noise \citep[using, e.g., a tool such as Trident;][]{trident-2017ApJ...847...59H}. For spectral observations of metal ions, the two approaches can ``meet in the middle'' by generating column densities of specific ions. Column densities can be reverse-modeled from spectral line features and forward modeled from bulk gas properties. While simulations could in theory directly model individual metal ion densities---thereby removing the need to forward model ionic column densities---this is often computationally intractable. It is already quite uncommon to track individual element densities \citep[the notable exception being EAGLE;][]{schaye2014} much less individual ion densities. Therefore, inferring elemental and ionic gas fractions must be part of the forward modeling process.

Connecting ionic absorption and gas properties through either forward or reverse modeling involves several assumptions. In recent years there has been considerable effort invested in characterizing the uncertainties that these assumptions introduce into reverse modeling. For instance,
to extract physical data from quasar absorption line spectra it is commonly assumed that each absorption feature originates from a single gas ``cloud'' or ``clump'' along the line of sight (LOS). 
Instead, \citet{haislmaier2020} showed that multiple line components may be needed to accurately capture the information contained in an absorption feature. Single component analysis is only capable of reproducing the average metallicity of multiphase structures \citep{marra2021} while multiple spatially distinct clumps can contribute to the same observed absorption feature if they overlap in line-of-sight velocity space \citep{marra2022}.

There has also been effort in quantifying how much uncertainty the UVB contributes to the reverse modeling of absorption spectra. Both \citet{gibson2022} and \citet{acharya2022} have studied how the extreme-UV (EUV) portion of the UVB ($\approx 1$--1000 Ryd) affects inferences of number density and metallicity. The former finds that hardening the EUV slope from $\alpha_\mathrm{EUV}= -2.0$ to $-1.4$ causes the derived metallicity to increase by an average of 3 dex. The latter found that number density can vary by a factor of 4-6.3 from high density ($\approx 10^{-3}\ \mathrm{cm^{-3}}$) to low density ($\approx 10^{-5}\ \mathrm{cm^{-3}}$) gas regions respectively, and metallicity varies from 2.2 to 3.3 from high to low-density gas. These differences are attributed to the normalization and shape of the different UVB models used in the study.

While it is likely not always true, for both forward and reverse modeling it is assumed that CGM gas is in ionization equilibrium; that is, that individual ions are in both collisional and photoionization equilibrium. This is because not assuming equilibrium proves to be very expensive, as evidenced by \citet{katz2022} in which the author implements an ODE solver (RAMSES-RTZ) on cosmological simulation data to evaluate a variety of non-equilibrium astrophysical environments that include the CGM. While the model does provide results that align very well with other chemical solvers such as CLOUDY \citep{smith2008}, the computational cost of these makes it very difficult to implement in other cosmological simulation software where the computational cost is already very high.

Regardless, any model accounting for photoionization (at equilibrium or not) requires that we understand the nature of the metagalactic ultraviolet background (UVB)---the background radiation originating from quasars and newly-formed stars. However, the UVB is challenging to model as it requires its own broad assumptions about extragalactic star formation rates and the distribution of quasars throughout the universe and across cosmic time. Indeed, the UVB itself is an abstraction representing the \textit{average} UV radiation field at a point in space far from any single galaxy. Because of these challenges, many models for the metagalactic UVB have been created over the past few decades. This makes the UVB a potential source of uncertainty in the models that connect spectral observations and their underlying gas.

Though understanding model uncertainty is a necessary process, it is important to reiterate that efforts have so far been focused on the implications for reverse modeling. Forward modeling from simulations is extremely valuable for comparing simulations and observations so it is
imperative that we quantify the uncertainty introduced by common, necessary assumptions to the forward modeling process. As a start to this effort, 
the goal of the work presented here is to quantify the variation in absorber column density introduced by uncertainties in the metagalactic ultraviolet background. We do this by repeatedly post-processing cosmological simulations from the FOGGIE project \citep{peeples2019, simons2020} with a pipeline that is identical except for the choice of metagalactic UVB. We then match clumps between each post-processing pass based on their physical location in order to compare the forward modeled column densities and underlying volumetric densities and temperatures. In Section \ref{sec:methods}, we discuss our data selection, simulation pipeline, and general methodology for performing our analysis, including the algorithm for matching clumps. In Section \ref{sec:results}, we present the results of our analysis and in Section 4 we interpret our findings and discuss the implications of our results in the broader research literature. Finally, we summarize our findings and discuss possibilities for future research projects in Section \ref{sec:conclusion}.

\section{Methodology} \label{sec:methods}

\begin{figure*}
    \centering
    \includegraphics[scale=0.5]{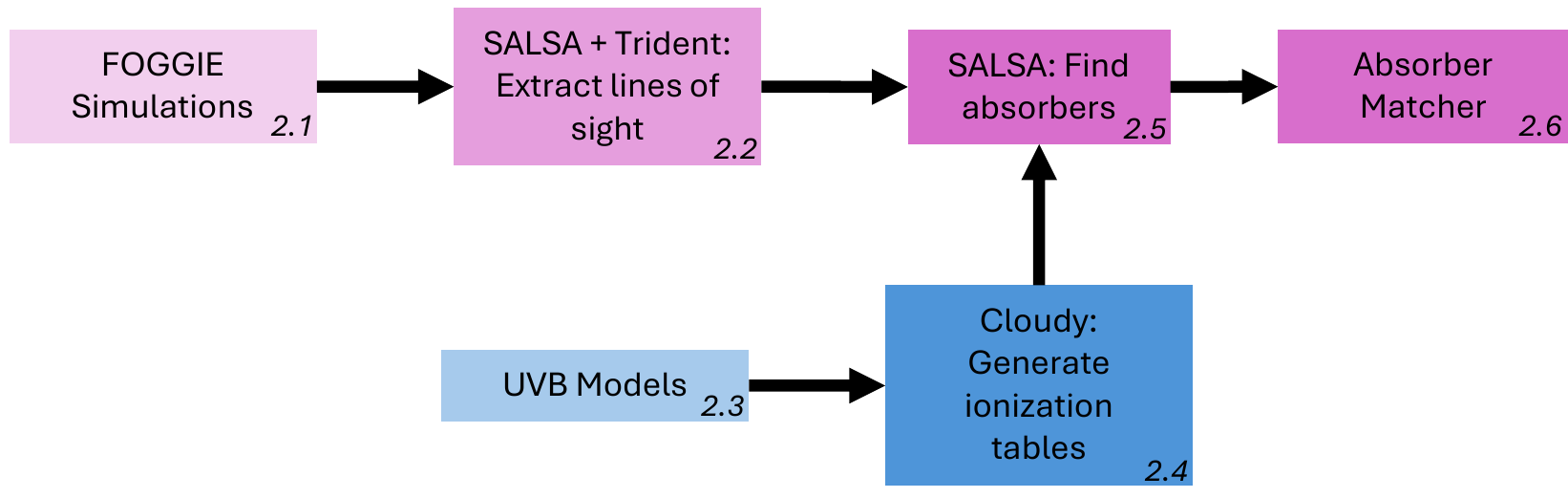}
    \caption{
    Overview of the analysis pipeline used in this work. Each block represents a different calculation and/or data product, as detailed in the text of this paper. In the bottom right corner of each box is the section in this work that each block corresponds to.  
    \label{fig:pipeline}}
\end{figure*}

For clarity, we present our process in the form of a flowchart that summarizes our data analysis pipeline in Figure \ref{fig:pipeline}. Each block of the figure corresponds to a subsection below. We base our analysis on the FOGGIE simulations (Section \ref{sec:foggie}). We use the Synthetic Absorption Line Surveyor Application \citep[SALSA; ][]{Salsa2020JOSS....5.2581B} to cast lines of sight or ``rays'' within a specified range of impact parameters (Section \ref{sec:extracting los}). SALSA is built on Trident, a synthetic absorption spectra tool \citep{hummels2019} that is used by SALSA to estimate ion column densities. Trident is itself built on yt \citep{yt_2011ApJS..192....9T}, which it uses to load the FOGGIE simulation data. The number densities inferred by Trident are  used by SALSA to find absorbers within each ray and calculate their column density and other properties. 

We adopt four different metagalactic ultraviolet backgrounds as outlined in Section \ref{sec:uvb} and use CLOUDY \citep{2013RMxAA..49..137F} to generate tables of ionization fractions (Section \ref{sec:cloudy}). Then using SALSA we find a set of absorbers for each (Section \ref{sec:absorbers}). For each set of SALSA absorbers created from each UVB, we match the absorbers based on their spatial position along a given line of sight (LOS) so that pairwise comparisons can be made (Section \ref{ssec:abs_cat}.)

\subsection{The FOGGIE Simulations}\label{sec:foggie}

The FOGGIE (Figuring Out Gas \& Galaxies In Enzo) Galaxies are a suite of cosmological zoom-in simulations of Milky Way-like galaxies that were made using the Enzo simulation code \citep{peeples2019, simons2020, brummel-smith19-ENZOAdaptiveMesh} \footnote{Also see the FOGGIE website at \url{https://foggie.science}} . One defining feature of the FOGGIE simulations is its high spatial and mass resolution in the circumgalactic medium, allowing for a more precise analysis of this region than in typical cosmological simulations. This is also the main reason we selected these galaxies for our study. We base our analysis on the ``Hurricane'' galaxy at redshift 2.5, around cosmic noon when the intensity of UV radiation is very high due to the high cosmic star formation rate density and thus when one might reasonably expect that the impact of variations in the ultraviolet background (UVB) might be largest. It should be noted that this galaxy was  simulated using the \citet{haart2012} UVB. While it may be more accurate to perform our analysis with multiple different sets of simulation data for each UVB, the aim of this work is not to analyze the variation of large-scale structures within the CGM, but to analyze the variation of ionization densities along sight lines. For this, performing post processing is sufficient for the scope of our project.
 
One limitation of the FOGGIE simulations is that they only follow the evolution of a single ``metal field,'' in which all of the CGM metals are tracked assuming constant abundance ratios (though variable absolute abundances), ignoring traces of individual species of ions. This then necessitates some form of post-processing method to extract data for these untracked ions.

\subsection{Extracting Lines of Sight}\label{sec:extracting los}

We use SALSA to mimic observations by generating a series of randomly oriented lines of sight, or ``rays,'' through the CGM. SALSA calculates random start and end points for our rays given a central point, ray length, and a range of impact parameters. This lets us approximate observational sightlines passing within a certain distance (i.e., impact parameter) of the galaxy center. 
he virial radius of the Hurricane galaxy is 41.70 kpc. The random sightlines are set to generate at a distance of 2.125 - 21.25 kpc from the galactic center. This ensures that these sightlines are passing primarily through the CGM and not the interstellar or intergalactic mediums.
In this way, we generate 100 rays through our simulated halo. These same 100 rays will be reanalyzed using different UV backgrounds (Section \ref{sec:uvb}).

With the spatial orientation of the rays specified by SALSA, Trident is used to infer the ionic number densities using CLOUDY-generated  equilibrium ion fraction tables (see Section \ref{sec:cloudy}) using the following equation: 
\begin{equation}\label{eqn:ion_fraction}
    n_{X_i} = n_Xf_{X_i},
\end{equation}
where $n_X$ is the total number density of the element as determined by the elemental abundances \citep[for this work we use Solar abundances; ][]{2016AJ....152...41P},   $f_{X_i}$ is the equilibrium ionization fraction of a given ion (which is a function of density $n$, temperature $T$ and redshift $z$), and $n_{X_i}$ represents the inferred ion number density \citep{hummels2018}.

\subsection{Ultraviolet Background Models}\label{sec:uvb}

\begin{figure}
    \centering
    \includegraphics[scale=0.5]{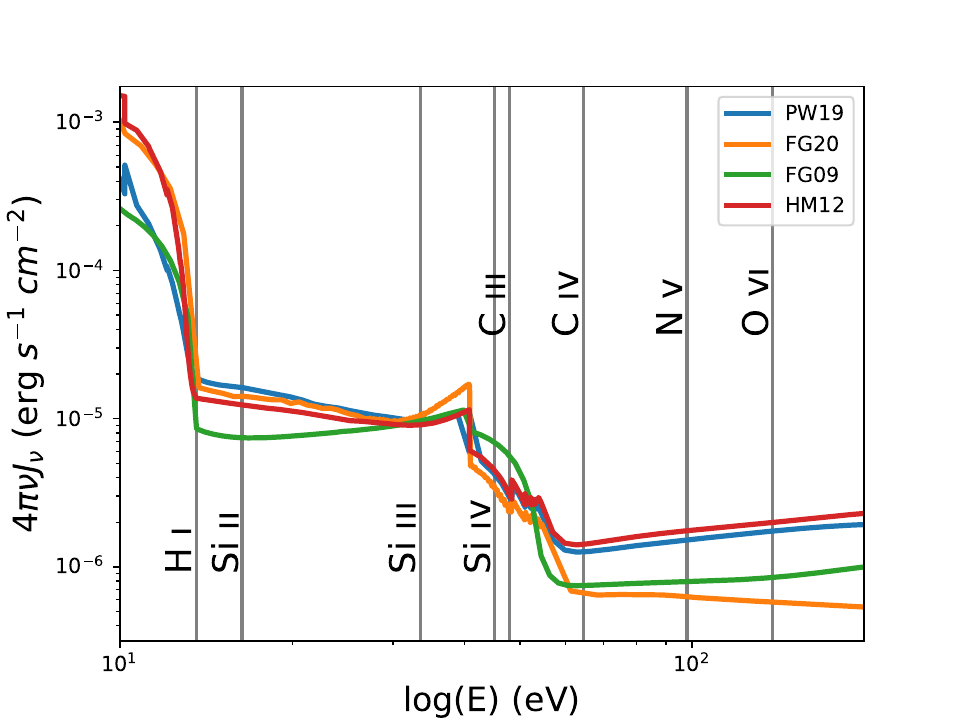}
    \caption{Radiant flux of the four ultraviolet backgrounds used in this comparison at $z=2.5$. The x-axis is the energy of the UVB in units of eV and the y-axis is the intensity of the UVB in units of erg s$^{-1}$ cm$^{-2}$.}
    \label{fig:uvb}
\end{figure}

We use two different ``families'' of metagalactic ultraviolet background (UVB) models with two models each for a total of four UVB models in total. These families are \citet{fauchergiguere2009} and \citet{fauchergiguere2020} (FG09 and FG20 respectively), and then \citet{puchwein2019} (PW19) and \citet{haart2012} (HM12). These models were selected both to allow us to analyze the differences between model families and the differences between generations of models within a single family. 
PW19 was developed in response to new photoionization and photoheating rates being determined by \citet{onorbe2017} as well as new calibrations to their effective rates to reach a higher optical depth ($\tau_e=0.065$) to match the Plank 2015 observations \citep{fauchergiguere2020}. In FG20, the authors make a number of updates based on updated galaxy UV luminosity functions, a new stellar spectral template, new AGN luminosity functions, improved IGM opacity measurements, updated Ly $\alpha$ forest constraints, Plank 2018 reionization constraints, and, finally, updated observational constraints on \heii\ reionization \citep{fauchergiguere2020, plank2016, planck2020}.

In Figure \ref{fig:uvb}, we show the intensity spectra (in units of erg s$^{-1}$ cm$^{-2}$) of all four UVBs that we use in this work. Each of the vertical black lines represents the ionization energies (in eV) of each of the ions used in this study: \hi, \siii, \siiii, \ciii, \siiv, \civ, \nv, \ovi. We include \siii\ to \siiv\ to evaluate the effect of the UVB for different ionization states of a single element. \ciii\ and \siiv] are in a region of the spectrum that is changing very rapidly and comparing these ions with similar ionization energy may help reveal the effects of the nonlinear portion of the spectra on column density. \civ, \nv, and \ovi\ are included due to their observational significance. \civ\ is useful for tracing CGM cold gas \citep{zheng2020, werk2016, lopez2024}. 

\subsection{CLOUDY Ionization Tables}\label{sec:cloudy}

As seen in Equation \ref{eqn:ion_fraction}, the UVB is coupled to the pipeline through equilibrium ionization fractions, $f_{X_i}$. We use CLOUDY to generate a table of these fractions for a broad range of densities and temperatures.
Both the FG09 and HM12 models have self-shielding ionization fraction tables available as part of the Trident project\footnote{https://trident-project.org/data/ion\_table/}. 
The newer FG20 and PW19 models do not have readily available tables made for them. To generate tables for these UVBs, we ran a series of one-zone, self-shielded CLOUDY models to determine the equilibrium ionization fractions for selected elements (see Section \ref{sec:uvb} for a list). These models were assembled into a table using the same code as the older FG09 and HM12 models
\footnote{https://github.com/brittonsmith/cloudy\_cooling\_tools}. 

\subsection{Absorber Extraction}\label{sec:absorbers}

Once Trident has inferred ionic number densities using the CLOUDY ionization fraction table corresponding to one of our four UVBs, SALSA is able to identify absorbers within the rays it randomly placed. SALSA does this iteratively for each individual ion using the Simple Procedure for Iterative Cloud Extraction (SPICE) method. A key assumption of this algorithm is that regions of high column density should give rise to detectable absorption features.

This algorithm works by setting an ion number density threshold above which lies some fraction of the ray's total column density of the ion in question. By default, the cutoff is 0.8. Then, bounds are set that define distinct ``clumps'' of gas that fall above this cutoff. That is, regions of space are identified that account for 80\% of the total column density of that ion. On the next pass, additional regions are flagged that account for 80\% of the column density that remains unaccounted for after the first pass. Clumps from each pass are combined if their average line-of-sight velocities are within 10~km/s of each other. This process is repeated until the column density of the remaining data that has not been assigned to an absorber is below the minimum density threshold. Therefore, the SPICE algorithm is controlled by two free parameters: the cutoff fraction and the minimum column density.

To ensure our results are not sensitive to these two free parameters, we employ a pseudo-grid search to determine the optimal set of parameters to apply to this algorithm. We track the distribution of $\log(N_\mathrm{H~I})$ for absorbers found by independently varying both parameters. The parameter not being analyzed is left at its default setting (0.8 for the cutoff fraction and $10^{13}\ \mathrm{cm^{-2}}$ for the minimum column density). We did not investigate the non-linear effects from varying both settings at once. Our aim is to find a range of settings for cutoff fraction and minimum density in which the $\log(N_\mathrm{H~I})$ distribution does not change significantly.

\begin{figure*}
    \centering
    \includegraphics[scale=0.45]{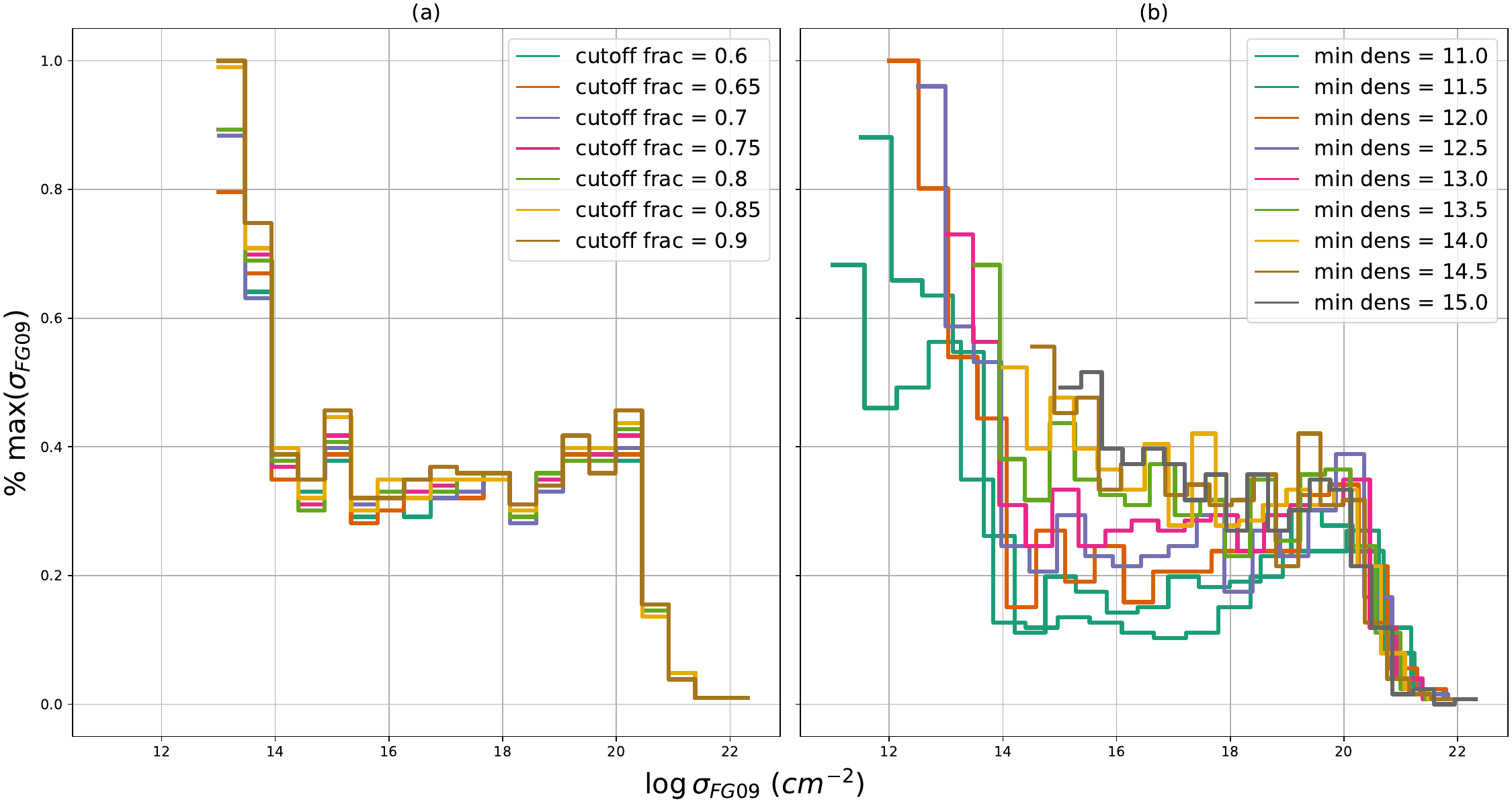}
    \caption{(a) A histogram showing the column density of SALSA absorbers produced using the SPICE method with cutoff fractions from 0.6--0.9 in intervals of 0.1. (b) A histogram of SALSA absorbers produced at different minimum density thresholds from $\log(N_{HI})= 11.0$--14.0 in intervals of $\log(N_{HI})=1$. It should be noted that both histograms are normalized to the highest column density in their respective figures for ease of comparison between parameter settings.}
    \label{fig:salsa_params}
\end{figure*}

In Figure \ref{fig:salsa_params} we plot a histogram showing \hi\ column densities of SALSA absorbers for two different SALSA settings: cutoff fraction (a) and minimum density (b). For our cutoff fraction distributions in panel (a), we select four different settings (including the default fraction of 0.8) from 0.6--0.9 in intervals of 0.1 while the minimum density cutoff remains at log$(N_{HI})=13.0$. In panel (b), the minimum column density cutoff is plotted at four different settings (including the default minimum column density setting of $10^{13}\ \mathrm{cm^{-2}}$) ranging from $\log(N_{HI})=11$--14 in intervals of $\log(N_{HI})=1$.

From our variation of the cutoff fraction in panel (a) of Figure \ref{fig:salsa_params}, we find that the overall distribution remains generally consistent despite the different settings used. Even though the total number of absorbers that SALSA identifies steadily decreases with increasing cutoff fraction, this decrease is small relative to the total number of absorbers found. This indicates that the cutoff fraction does not have a significant impact on the results of our analysis, so we elected to leave this parameter at its default value of 0.8. 
In panel (b) of Figure \ref{fig:salsa_params} we see that as the minimum column density parameter increases the total number of absorbers detected decreases rapidly, with the overall \hi\ column density distribution skewing towards higher column densities. Unlike with the cutoff fraction, the \hi\ column density distribution does not remain stable as we vary the minimum column density. Instead, we chose to adopt different minimum column densities for each ion of interest. For \hi, \ciii, \civ, and \ovi\ we use $\log(N) = 12.5$. For \siii\ and \siiii, we use $\log(N_{\mathrm {HI}}) = 11.5$. Finally, for \nv, we use $\log(N_{HI}) = 13.0$. We select these values based on the minimum column density these absorbers could potentially be detected by observation \citep{tumlinson2011, tumlinson2013, werk2013, werk2016, bordoloi2014, bordoloi2018, lehner2011, lehner2022}.

Once SALSA has identified absorbers for each ion based on their number density it reports gas properties such as volumetric density, temperature, and metallicity for each absorber. These quantities are a column density-weighted average of the cells belonging to that absorber.

\subsection{Absorber Matching} \label{ssec:abs_cat}

The UVB will affect the overall ion number densities of the ray and therefore the clumps identified by the iterative SPICE algorithm as the algorithm uses number density cutoffs. Once the SPICE algorithm has been run on our 100 rays for each of our four UVBs, we must match absorbers based on their position along each LOS. This is because our analysis relies on pairwise comparison between the same absorbers from different UVBs. Altering the UVB and overall ion number density can change the shape of the identified absorbers, so we categorize them into different groups based on their relative size and position along a given ray. The categories are as follows:

\begin{enumerate}
    \item Match: absorbers are of the exact same size and position along the ray, covering the same resolution elements in the underlying cosmological simulations.
    \item Different Size: the two absorbers are different sizes (i.e., they span a differing number of underlying simulation cells) but match in either start or end position. In other words, one UVB results in a clump that is longer or shorter than that from another UVB, but they still occupy the same physical region along the ray.
    \item Overlap: absorbers have a significant overlap with one another along the LOS, but they do not fully line up in terms of size or position (i.e., they do not share a start or end point and thus are likely to encompass a different number of cells).
    \item Merge: when one of the UVBs produces multiple small absorbers but the other UVB produces only one large absorber that is a superset of the smaller ones and spans the same spatial extent.
    \item Lonely: there exists an absorber in one UVB while the other UVB has no absorber.
\end{enumerate}

\begin{figure}[!h]
    \centering
    \includegraphics[scale=0.4]{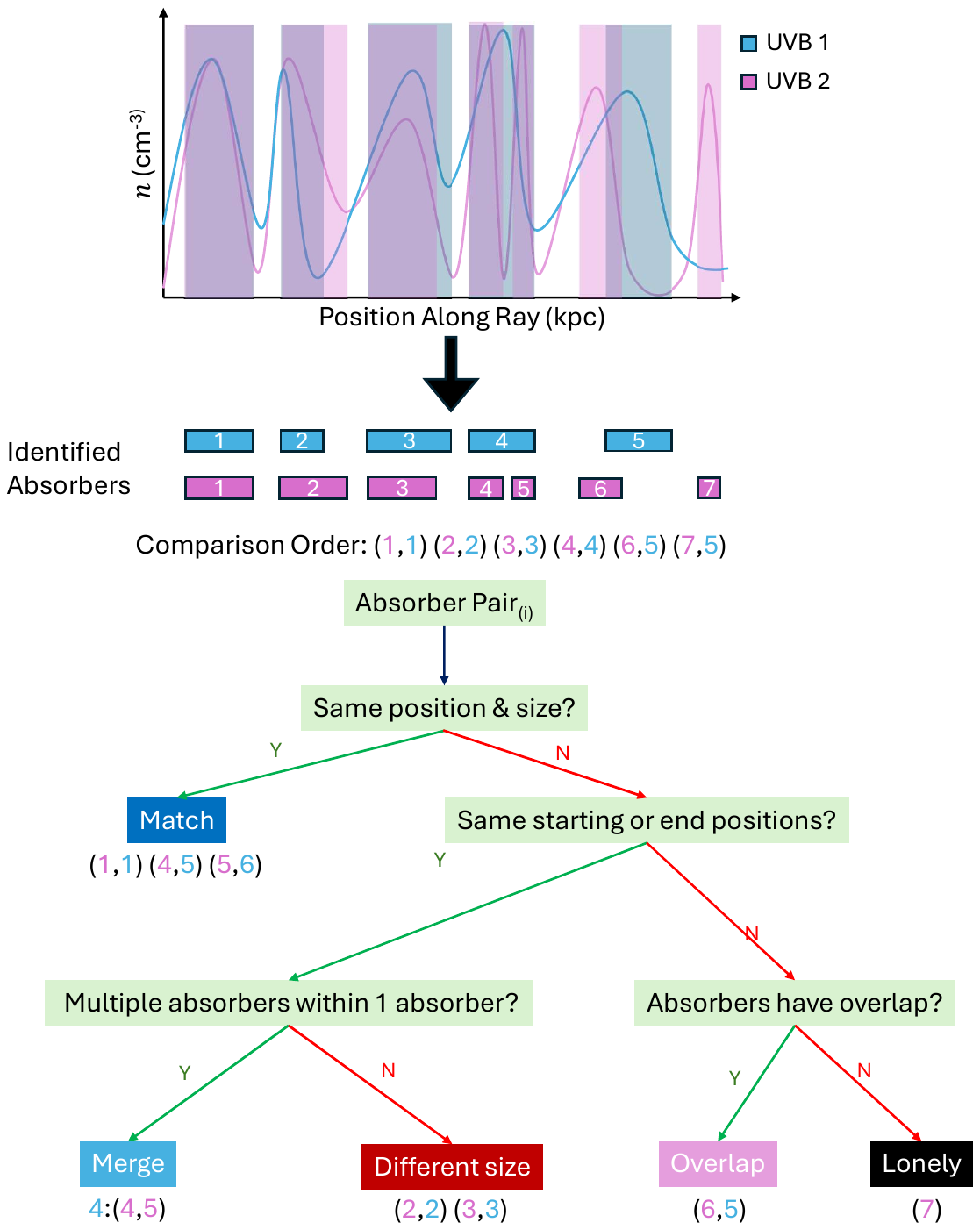}
    \caption{Flow chart of the pairwise comparison algorithm. Each absorber is numerically labeled in the order they appear along the ray. The ``comparison order'' list indicates the order in which pairs of absorbers are put through the flow chart.}
    \label{fig:flowchart}
\end{figure}

All of these comparisons are made using a series of Boolean comparisons with some margin of error allowed. This margin of error prevents situations in which, for example, a pair of absorbers would fall into the Different Size category when the size of the two absorbers has only a one-cell difference. 

Specifying a margin of error in terms of cells is difficult for this analysis as the ray can have a variable path length $dl$ through each cell in the underlying simulation, even if those cells have uniform spatial resolution. To address this issue we select a margin of error based on a fraction of the median size of absorbers. Through our analysis pipeline, we find that each UVB has a median absorber size of 28 cells except for PW19 which is 23 cells long. Choosing to select 28 cells given the majority of UVBs have this as their median absorber size, we select a margin of error roughly $\pm 5\%$ of this value that equates to 1.4 cells, which we rounded up to 2 cells.


In order to compare absorber properties across UVBs, the small individual clumps in Merge cases had their properties combined. That is to say, the column densities of these absorbers were linearly summed together, while other physical quantities of the absorber, such as temperature and gas density (see Section \ref{sec:absorbers}), were combined by weighted average based on the column densities of the individual absorbers using the following equation: 
$$\overline{x} = \frac{\sum_{i=1}^k (x_i \cdot N_{i,abs})}{\sum_{i=1}^{k}N_{i,abs}}$$
where $\overline{x}$ is the weighted average quantity, $k$ is the total number of simulation cells spanning the merging absorbers, $x_i$ is the given physical quantity in cell $i$, $N_{i,abs}$ is the column density for a given absorber in cell $i$.

Lonely cases were removed from the analysis as they have no partner to compare to. This indicates that the one of the two UVB absorbers either does not exist at that position along the ray or it is below the detection threshold for that ion (Section \ref{sec:absorbers}). The fraction of lonely absorbers is $\simeq 1\%$ for each pairwise comparison (i.e., for each pair of UVBs that are compared).

Additionally, there were a few rays for which the sorting algorithm was unable to categorize absorber pairs. These cases consisted of absorbers that fell into multiple categories, or rays that had no absorbers.
Instead of attempting to develop categories for all of these outliers, the rays are removed from the analysis. The number of removed rays tends to stay between 0--10\% for most ions. However for \nv, 37--68\% are removed from the analysis depending on which pair of UVBs are compared. This is largely due to the fact that the abundances of this nitrogen ion are significantly lower than the other ions used in this work. Thus, SALSA recognized many of the \nv\ rays as having no observable absorbers.

\begin{deluxetable}{ccccc}
\tablewidth{5pt}
\tablecaption{Total SALSA Absorber Counts \label{tab:tot_abs_counts}}
\tablehead{
\colhead{} & \colhead{FG09}& \colhead{FG20} & \colhead{HM12} & \colhead{PW19} \\
}
\startdata 
       \hi & 618  & 618  & 618  & 618  \\
     \siii & 442  & 436  & 423  & 577  \\
    \siiii & 573  & 589  & 540  & 688  \\
     \siiv & 468  & 487  & 451  & 519  \\
     \ciii & 608  & 623  & 588  & 644  \\
      \civ & 561  & 548  & 544  & 594  \\
       \nv &  37  &  30  &  85  & 129  \\
      \ovi & 523  & 468  & 562  & 622  \\
\enddata
\caption{Shows the total number of absorbers SALSA found per ion for each UVB before any ray / absorber removal is performed on the data.}
\end{deluxetable}


Table \ref{tab:tot_abs_counts} shows the number of absorbers identified by SALSA for each UVB. It should be noted that this is before pairwise comparison is performed on the data and rays are removed from the data. Assuming a roughly equal number of absorbers per ray, this means that the actual number of absorbers is 0--10\% lower than the listed absorber count. Excluding \nv, this leaves us with at minimum approximately 380--556 absorbers per UVB for comparison, which is a sufficiently large number for the purpose of this study.

\section{Results} \label{sec:results}

Using the methodology described in Section \ref{sec:methods} we make three pairwise comparisons: FG09 and FG20 (Faucher-Gigu\`ere family), HM12 and PW19 (Haardt-Madau-Puchwein family), and finally FG20 and PW19, the latest models from each family. We begin by comparing the total column densities along each ray and then perform a comparison between individual absorbers and their physical properties as identified by the pipeline described in Section \ref{sec:methods}. It should be noted that for this section, we will be making references to various forms of column density and gas density. For the sake of clarity, we will be defining each of these terms here: $N_\mathrm{abs}$ refers to absorber column density ($\mathrm{cm^{-2}}$), $N_{\mathrm{LOS}}$ indicates the total column density along the entire LOS ($\mathrm{cm^{-2}}$), and  $n$ is the gas density of a given absorber ($\mathrm{cm^{-3}}$).

It should also be noted that in addition to examining ion number density and temperature differences for each species we also looked for trends relating to variation in metallicity.  We did not find any discernible trends relating to metallicity for any of the ions considered in this work, and thus for the sake of clarity we have not showed those results in the figures presented below.

\subsection{Total Column Density Comparison}

\begin{figure*}
    \centering
    \includegraphics[scale=0.28]{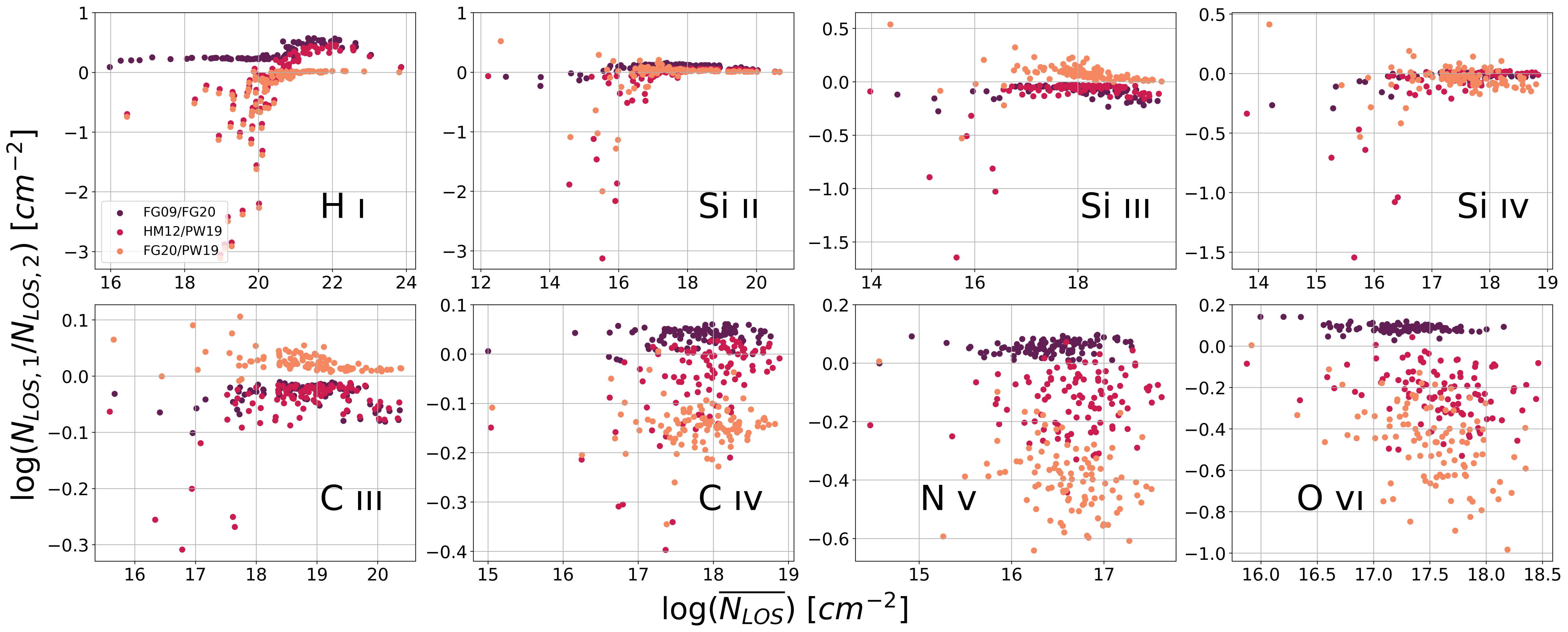}
    \caption{Differences in total column density along the LOS as different UVBs are applied. Differences are quantified as  $\log(N_{LOS,1}/N_{LOS,2})$ (with "1" and "2" being the UVBs defined in the legend), the ratio between new and old UVBs as listed in the figure legend. Quantities on the x-axis are the average of the total column densities between the two UVBs in the comparison (written $\log(\bar{N})$). The points in purple refer to the comparison between FG20 and FG09 ($\log{\mathrm{FG09/FG20}}$),  the magenta points refer to PW19 and HM12 ($\log{\mathrm{HM12/PW19}}$), and the orange to PW19 and FG20 ($\log{\mathrm{FG20/PW19}}$). The different ions used in the comparison are ordered in terms of increasing ionization energies from left to right and top to bottom. It should also be noted that the x and y axis of each subfigure is on a different scale. }
    \label{fig:tot_col_dens}
\end{figure*}

Figure \ref{fig:tot_col_dens} compares the total column densities between each of the three model pairs for each ion considered in this work, plotting the average $\overline{N_\mathrm{LOS}}$ in the x-axis, along with the $\log$ difference between the two UVBs in the y-axis. The points in purple refer to the comparison between FG20 and FG09 ($\log{\mathrm{FG20/FG09}}$),  the magenta points refer to PW19 and HM12 ($\log{\mathrm{PW19/HM12}}$), and the orange to PW19 and FG20 ($\log{\mathrm{FG20/PW19}}$). The different ions used in the comparison are ordered in terms of increasing ionization potential from left to right.

Throughout each comparison we find that the FG20 and FG09 models tend to have the best agreement, with differences between the two UVBs never being larger than $\simeq 0.75$ dex for any ion. These differences also tended to be consistent for the entire density range. The scatter in log difference generally remaining positive or negative depending on the ion, reflecting systematic differences in the species densities resulting from the two UVBs. On the other hand, the PW19/FG20 comparison has the largest differences. 

Looking at the differences between pairwise comparisons of total ion column densities, we find that the differences between UVBs display distinct trends.
The total \hi\ column densities have notable systematic offsets (with a typical offset of $\simeq 0.3$ dex between pairs of models), with very large differences in total column densities between compared sight lines for a narrow range of total column densities. 
 \siii, \siii, \ciii, and \siiv\ have a very strong agreement between their distributions with very small typical offsets between UVBs (generally $\simeq 0.1$ dex), although a small fraction of the sightlines (particularly those with low total column densities) have notable differences. 
\civ, \nv, and \ovi\ display significant offsets between models (particularly when comparing the PW/HM vs. FG families of models) as well as significant scatter within those offsets.  This offset and scatter generally grows with increasing ionization potential.

Starting in the \hi\ panel (top left), in the region between $\log\overline{N_\mathrm{LOS}} = 19$--20 the amount of \hi\ in PW19 rays is significantly larger than in the HM12 rays by up to a factor of $\simeq 10^3$. Below this range of $\log\overline{N_\mathrm{LOS}}$, the column densities have good agreement. In the \siii\ panel there is similar behavior from $log\overline{N} = 14$--16, though less pronounced than observed in \hi. In \siiii and \siiv we see PW19 has significantly higher column densities than HM12 at $\log\overline{N_\mathrm{LOS}} < 16.5$. FG20 and PW19 appear to have agree well for \siiii and \siiv with $|\log{\mathrm{FG20/PW19}}| < 0.5$ for almost all absorbers.

\subsection{Pairwise Comparison Summary}

\begin{figure*}
    \centering
    \includegraphics[scale=0.4]{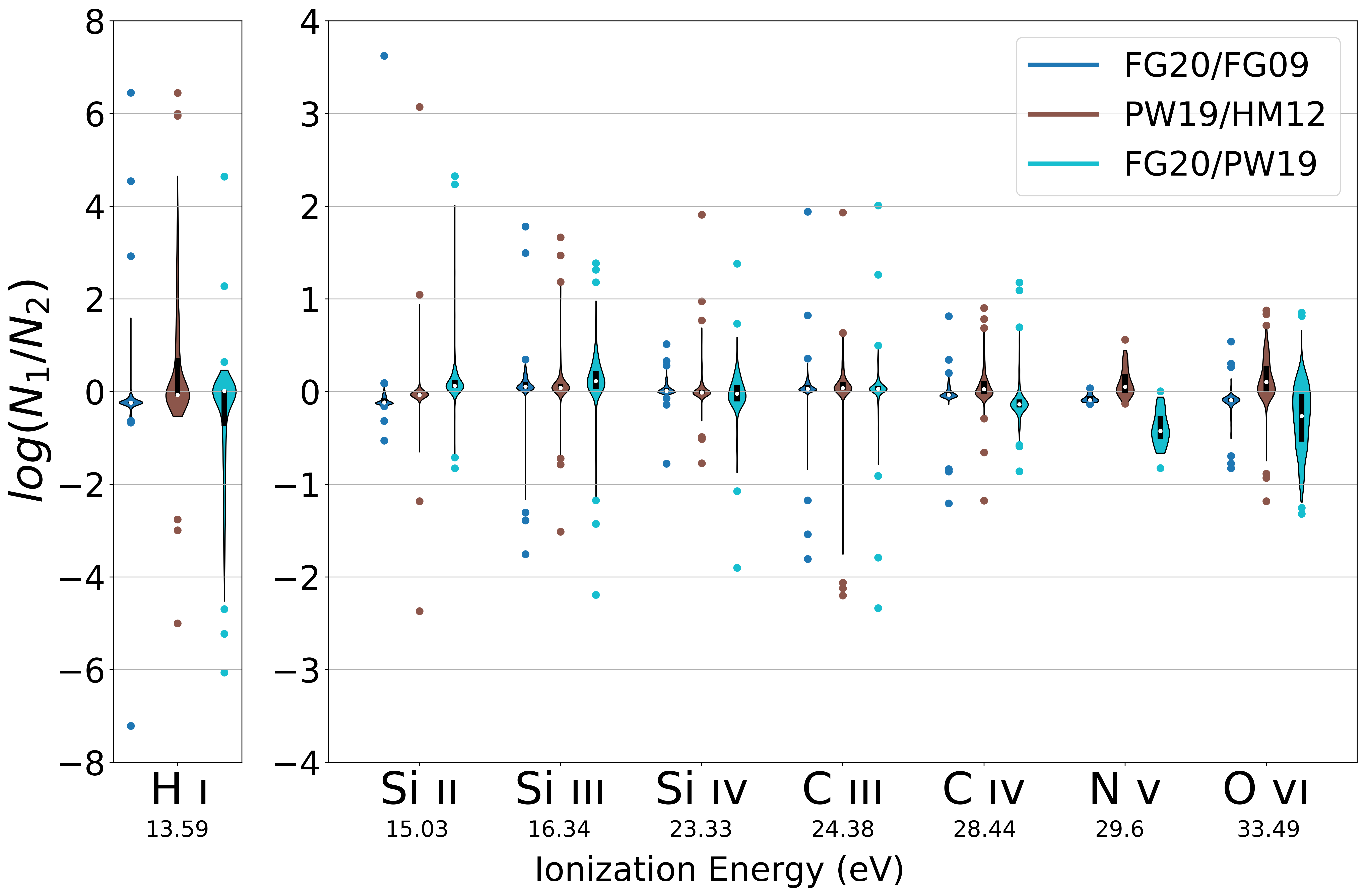}
    \caption{Violin plots showing the column density differences through each of the UVB pairwise comparisons for each ion. Along the vertical axis is the ratio between the two backgrounds, and the x-axis is the ion being analyzed, ordered in terms of increasing ionization energy from left to right. Each ion has three different distributions showing the ratio of $N_1$ to $N_2$, the models by which they originate being indicated in the legend of the figure. Within each violin plot, the white dot represents the median value of the distribution and the black bar represents the inner-quartile range of the data. For the sake of clarity, \hi\ has been plotted on a separate axis as it has a much broader variation than the other ions.}
    \label{fig:summary}
\end{figure*}

Figure \ref{fig:summary} is a series of violin plots that indicate the distribution of column density differences for each UVB pairwise comparison for each ion used in our analysis. The y-axis is the column density difference shown as a ratio of $N_1/N_2$ where $N_1$ and $N_2$ represent the absorber column densities for models indicated in the legend of the figure. The x-axis shows the ions along with their ionization energies listed in order of increasing ionization energy. In each violin plot, the white dot near the center represents the median column density difference and the black bar is the inner quartile range. The dots on the edge of each plot are outliers (that is to say, data that falls outside of the 99th percentile).

Here, we see that the \hi\ distribution is significantly different than the other ions, covering a much larger range of values with a much wider spread in the data. As such, it is plotted on a separate axis from the other ions to ensure that features relating to the other ions are still legible. For the rest of the ions, we see that for \siii, \siiii, \siiv, \ciii\ and \civ\ the distribution of ions tends to remain fairly consistent, with most ions having median differences $\approx 0$. For higher-ionization-energy ions, the distribution of ions become much more spread out with inner quartile ranges that extend to $\approx 0.5$ dex as opposed to the inner quartile ranges of lower-ionization-energy ions, which only reach inner quartile ranges of $\approx 0.2$ dex. Additionally, the number of outliers is very small, with any potential outliers that remain within 0.5 dex of the distribution unlike many of the lower-ionization ions. 

\subsection{One-to-One Comparisons}

\begin{figure*}
    \centering
    \includegraphics[scale=0.4]{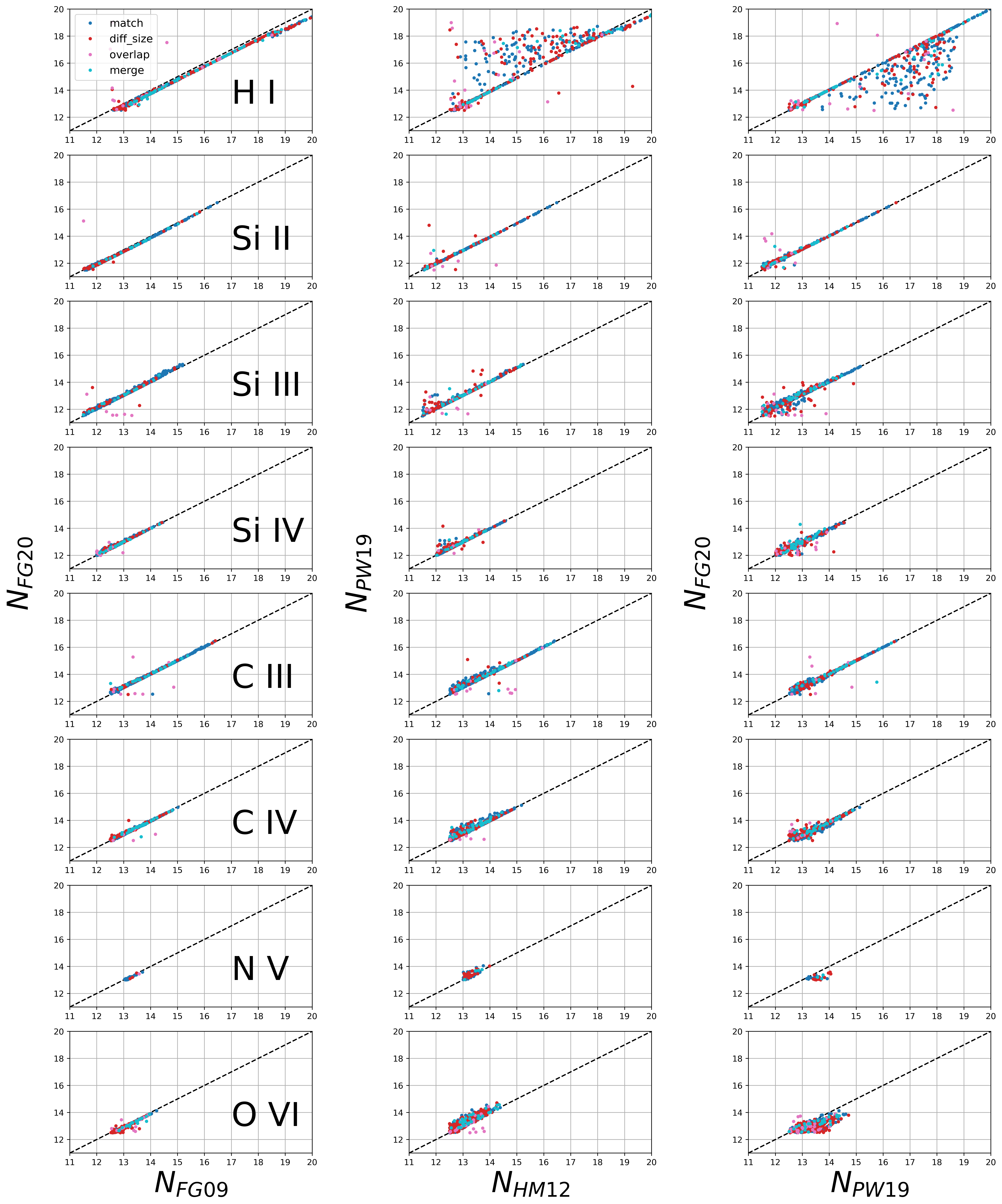}
    \caption{Direct comparison of absorber column density for each model pair and each ion in the analysis. The x-axis contains the log column density for absorbers in the older UVB models, while the y-axis contains log column densities of the newer models. The dashed black line at the center of the figure indicates the line of equality between the to models. The pairwise comparisons going from left to right are: FG20 vs FG09, PW19 vs HM12 and FG20 vs PW19.
    }
    \label{fig:1:1comparison}
\end{figure*}

Figure~\ref{fig:1:1comparison} shows a direct comparison of absorber column densities for each model pair. Each row is of a different CGM ion listed from top to bottom in order of increasing ionization energy, while each column represents a different pairwise comparison. From left to right, the comparisons are as follows: FG20 in the y-axis and FG09 in the x-axis, PW19 in y-axis and HM12 in x-axis, FG20 in y-axis and PW19 in x-axis. Each point on the figure is colored based on which category the absorber comparison was sorted into from our absorber categorization (Section~\ref{ssec:abs_cat}). 

From this figure we see that below $\log{N_\mathrm{abs}} < 18$ in \hi\, the PW19 absorbers have larger densities than both HM12 and FG20. \siii\ has good agreements between models. From \siiii\ to \civ, the FG09--FG20 shows that the two UVBs create absorbers that are very similar to one another with very few outliers. HM12--PW19 in this same ion range show some variation between absorbers at $\log{N_\mathrm{abs}} < 14.5$ that grows more significant with increasing ionization energy (with the exception of \hi). Similar to the total column density results, PW19--FG20 has the largest differences between models in this range with the largest variation existing at $\log{N_\mathrm{abs}} < 14.0$ (again, with the exception of \hi).

The number of \nv\ absorbers is very sparse (Section \ref{ssec:abs_cat}) and thus it is very difficult to extract any significant trends from the data. However, it should be noted that all PW19 absorbers have higher column densities than FG20. For \ovi\ absorbers, FG09--FG20 shows that there is relatively little scatter between the UVBs, however there is a very clear systematic offset where FG09 absorbers are typically larger by around 0.1 dex. HM12--PW19 shows a large scatter with the majority of the PW19 ions having higher densities than HM12 across the full range of column densities. Finally, PW19--FG20 shows large scatter that is very similar to HM12--PW19 with PW19 absorbers having significantly higher column densities, with the main difference being that the difference between FG20 and PW19 absorbers grows larger with increasing column density while HM12--PW19 remains at a flat offset.

\subsection{FG09-FG20 Comparison}

\begin{figure*}
    \centering
    \includegraphics[scale=0.4]{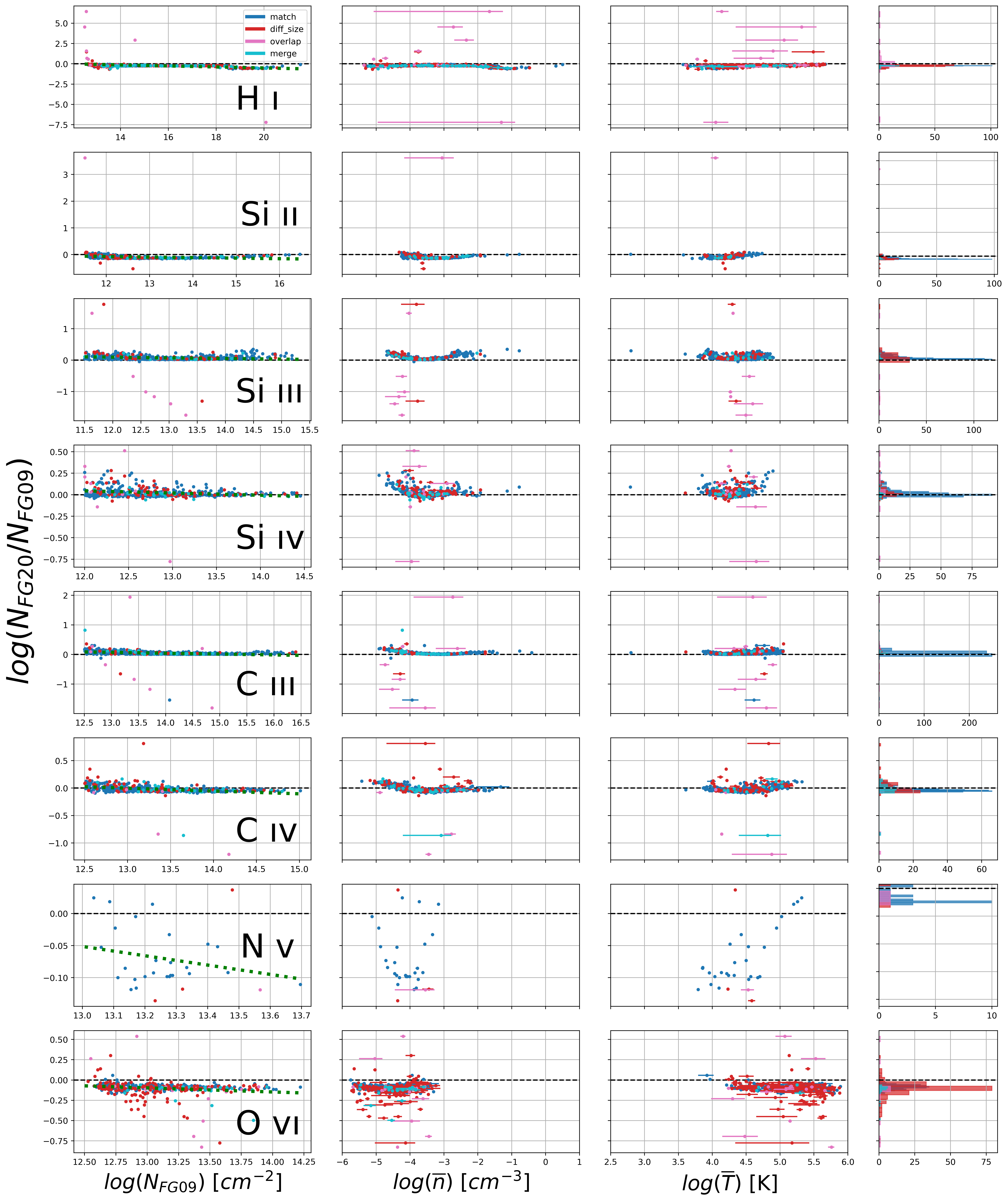}
    \caption{Difference in absorber column density between the FG 2009 and FG 2020 UVB models based on their column density, temperature, and gas density. The y-axis is the $\log_{10}$ ratio between FG20 and FG09. The leftmost column shows this ratio versus the FG09 absorber column densities. A dashed black line is included to represent the ``match'' line where the two absorber column densities match with one another. Additionally, a green dotted line is included as a linear fit generated to the comparison scatter. The second column shows the difference versus gas density: the x-axis gives the average gas density $\bar{n}$ between the two UVB models with error bars showing the range between individual absorber densities. The third column is versus temperature where the x-axis is again the average between the two models with error bars showing the smaller and larger values. The final column is a histogram showing the distribution of the FG20--FG09 difference ratio. Each row is a different ion used in the analysis and each point is colored based on their spatial relation to one another as outlined in Section~\ref{ssec:abs_cat}.
    }
    \label{fig:fg_v_fg}
\end{figure*}

Figure \ref{fig:fg_v_fg} shows the column density differences between individual FG09 and FG20 absorbers found using the SALSA algorithm and their relationship with different physical quantities in the simulation data. The column density difference is $\log{N}_\mathrm{abs\,FG20}$/$N_\mathrm{abs\,FG09}$ where $N_\mathrm{abs}$ represents the column density of a given absorber. In order from left to right, the comparison quantities are: FG09 absorber column densities, absorber gas density, temperature, and finally a histogram of the column density differences. Each row represents a different ion as indicated by the label in the panels in the leftmost column, listed in order of increasing species ionization potential.

In \hi, we see that ions using the FG20 UVB have systematically lower column densities than with FG09. Notably, at gas number density $\log{n} = -1$, the difference between the two UVBs increases to a maximum of $\sim 0.5$ dex. This difference appears to be associated with lower temperature absorbers ($\sim \log(T/K) =$4.0--4.5 K) as can be seen in the third column. In fact, most of the \hi\ absorbers with larger FG09 column densities tend to be associated with temperatures lower than $~\log(T/K) =$5 K.
\siii, \siiii, \ciii, \siiv\ and \civ\ all have very similar trends to one another.  At  $\log{\overline{n}} \leq -4$, FG20 tends to have higher densities that follow a downward trend until $\sim \log{\overline{n}} \simeq -4$, after which the UVB absorbers have similar column densities. \siii, while mostly following this same trend, appears to be shifted downwards by ~0.125 dex. In some of the other lower ionization-state ions, \siiii\ and \ciii, we see that around $\log{\overline{n}} = -3$ there is an upward trend where FG20 has higher column densities. The density difference remains $\simeq 0$ in \siiv\ and \civ. There are no noticeable trends with temperature except for \civ, where FG20 tends to have higher densities at $\geq \log\overline{T} = 4.625$
There are very few \nv\ absorption systems and thus it is very difficult to extract a trend from this species. It does appear, however, that the absorbers SALSA does identify have very similar column densities to one another. There is a faint trend with $\overline{T}$ where $ \geq \log\overline{T} = 4.6$ tends to have more similar column densities as temperature increases.
Finally, \ovi\ has no trends with gas density or temperature, but the FG09 UVB does have systematically higher column densities by a factor of $\simeq 0.1$ dex.

\subsection{HM12-PW19 Comparison}

\begin{figure*}
    \centering
    \includegraphics[scale=0.4]{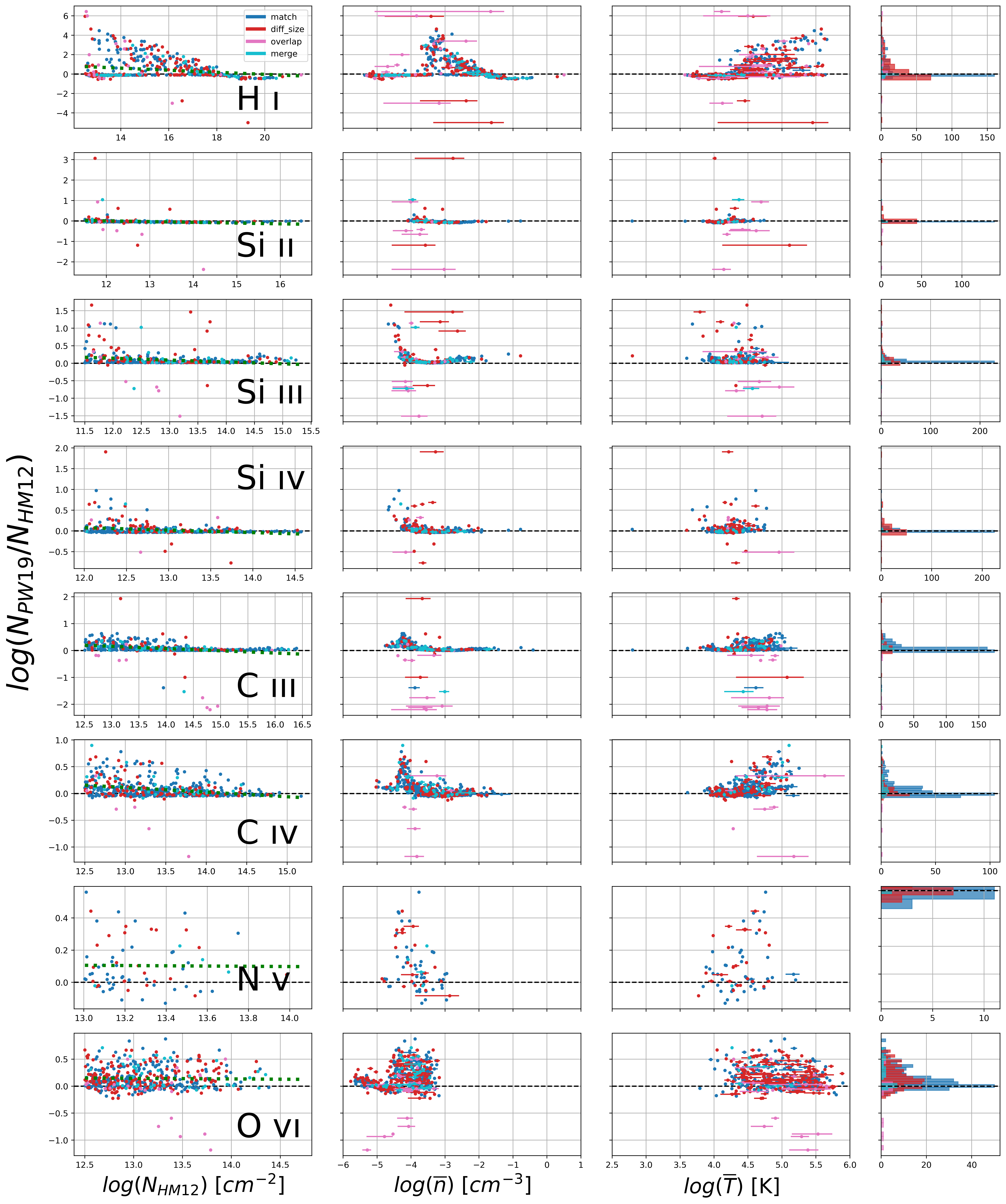}
    \caption{Similar to Fig. \ref{fig:fg_v_fg}, now comparing HM12 and PW19. The ratio between the two is $N_\mathrm{abs,\,PW19}/N_\mathrm{abs,\,HM12}$.}
    \label{fig:hm_v_p}
\end{figure*}

Fig. \ref{fig:hm_v_p} is a very similar plot to Fig. \ref{fig:fg_v_fg}, now with the comparison being between PW19 and HM12 and with the column density ratio being $N_\mathrm{abs,\,PW19}/N_\mathrm{abs,\,HM12}$. 

For \hi, we see from $\log{\overline{n}} = -4$ to $-1.5$ that there are significant differences between the two UVBs.  While absorbers typically have similar column densities (with better agreement at higher column density) there is significant scatter at relatively low column densities,  with a moderate but notable fraction of PW19 absorbers having column densities larger than their HM12 counterparts by several dex. Based on columns two and three, these differences are most pronounced in absorbers with intermediate densities and high temperatures ($\log\overline{T} > 4.25$ and $-4.5 < \log\overline{n} < -3.5$).
Excluding \siii, which does not appear to have many significant differences between the two UVBs, the Si and C ions both show significant column density differences at $\log{\overline{n}} < -4$. At higher densities, the column density differences tend to stay around $0$. While Si ion column density differences become less significant with increasing $\overline{n}$, C ion column densities with $\log\overline{n} = -4.25$ have larger column density differences peaking at ~$0.5$ dex for \ciii\ and ~$0.75$ dex for \civ.
As with Fig. \ref{fig:fg_v_fg}, there are relatively few \nv\ absorbers and thus it is very difficult to extract any useful trend from these panels.
\ovi, however, displays a notable trend with gas density. Above $\log{\overline{n}} = -4.5$, the column density difference immediately scatters with many PW19 absorbers having significantly larger differences that range from $0$ to $0.6$ dex. There are no notable trends with HM12 column density or $\overline{T}$.

\subsection{PW19-FG20 Comparison}

\begin{figure*}
    \centering
    \includegraphics[scale=0.4]{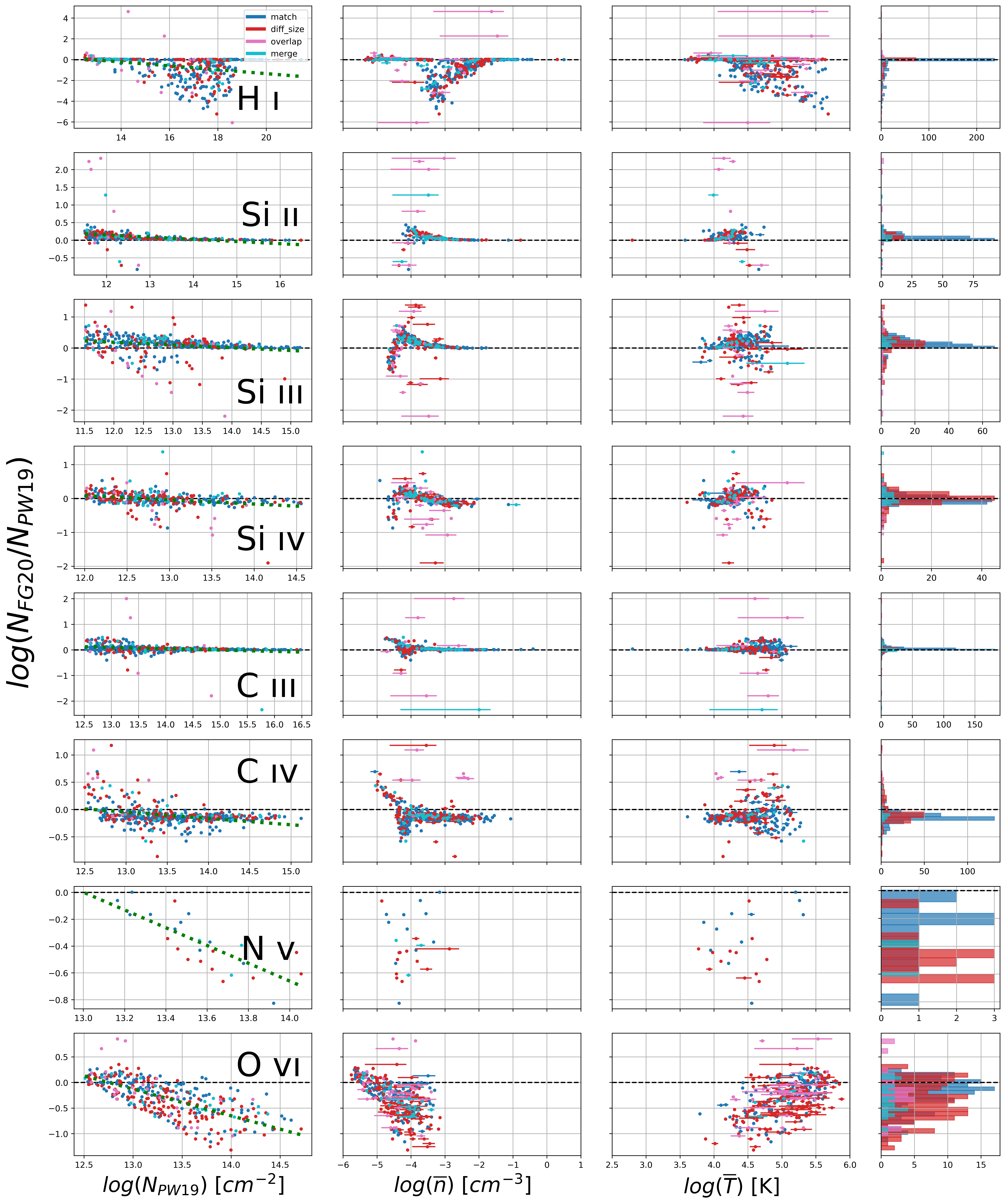}
    \caption{Similar to Fig. \ref{fig:fg_v_fg} and Fig. \ref{fig:hm_v_p}, now comparing FG20 and PW19. The ratio between the two is $N_\mathrm{abs\,PW19}$/$N_\mathrm{abs\,FG20}$.}
    \label{fig:fg_vs_p}
\end{figure*}

Fig. \ref{fig:fg_vs_p} is similar to the previous two figures, but now comparing PW19 and FG20 with a column density ratio of $N_\mathrm{abs\,PW19}$/$N_\mathrm{abs\,FG20}$. 

The \hi\ absorbers show a similar feature as seen in Fig. \ref{fig:hm_v_p} from $\log{\overline{n}} = -4$ to $-1.5$, with a maximum difference of approximately 4 dex. This structure is also reflected in Column 3, where the absorbers with larger column density differences generally exist at $\log\overline{T} > 4.25$.
\siiii, \ciii, \siiv\ and \civ, as in Fig. \ref{fig:hm_v_p}, have very little column density differences except for $\log{\overline{n}} < -4$ where the scatter becomes significantly larger. As with previous comparisons, the number of \nv\ absorbers are low and thus trends are difficult to discern. \ovi\ has overall very high scatter between the two UVBs, but it does appear that as gas density increases the column density difference grows larger in favor of PW19.

It should also be noted that in addition to examining ion number density and temperature differences for each species we also looked for trends relating to variation in metallicity.  We did not find any discernible trends relating to metallicity for any of the ions considered in this work, and thus for the sake of clarity we have not showed those results in the figures presented here.

\section{Discussion} \label{sec:discussion}

In this experiment we have used Trident and SALSA to create and analyze 100 randomly-oriented lines of sight through the circumgalactic medium of a simulated Milky Way progenitor galaxy from the FOGGIE project. We use these softwares to extract absorbers from each ray assuming four different metagalactic ultraviolet backgrounds from two different families of models (Faucher-Gigu\`ere 2009, Faucher-Gigu\`ere 2020, Haart \& Madauu 2012 and Puchwein et al. 2019). We then performed three pairwise comparisons (FG09 to FG20, HM12 to PW19 and PW19 to FG20) examining the evolution of these model families as well as comparison between the most current version of these models for eight different ions (\hi, \siii, \siiii, \ciii, \siiv, \civ, \nv, and \ovi). We examined differences in column density for entire lines of sight as well as individual absorbers, with a particular focus on the differences relating to the physical conditions of the absorbers.

\subsection{Interpretation of results}

Many---but not all---of the results presented in Section~\ref{sec:results} can be understood through differences in the overall magnitude of the metagalactic ultraviolet background at the ionization energies corresponding to the ions under consideration, as shown in Figure~\ref{fig:uvb}.  Figure~\ref{fig:tot_col_dens} shows ray-by-ray comparison of the total column density of each ion.  In general, species where the magnitude of the UVB flux varies significantly near their ionization energy (such as \hi, \nv, or \ovi) show significant offsets between the total amount of each ion, whereas species where the UVB flux is much closer at their ionization energy (e.g., \siiii) tend to show correspondingly smaller systematic offsets. 
Within these results, however, some lines of sight have extremely large column density differences (up to a few orders of magnitude for \hi\ and \siii\ over a narrow range of total column density).  Absorber-by-absorber comparisons (in Figures~\ref{fig:fg_v_fg}--\ref{fig:fg_vs_p}) suggest that this has to do with differences at relatively specific ranges of species gas density and temperature, which require further examination.  It is possible that this is because each species that we consider reaches its equilibrium by a combination of UV background-driven photoionization (both into and out of the ion in question) and collisional processes (both ionization and recombination). These processes depend nonlinearly on a combination of gas density and temperature as well as the properties of the metagalactic UV background, and these complex interactions must be the cause of the differences that we see. 

For the SALSA-extracted absorber pairs (Figures~\ref{fig:fg_v_fg}--\ref{fig:fg_vs_p}), we find that the trends in our results agree with the combination of photoionization and collisional reactions that control the ion balance at a given point in thermodynamic phase space.
The scatter in column density ratio increases with increasing species ionization energy, which is reasonable given that the metagalactic ionizing background is relatively poorly constrained at those high energies and thus the models vary substantially in their predictions for the UVB flux.  This translates into significant differences in the region of temperature-density phase space where one would expect photoionization processes to dominate over collisional processes---as the magnitude of the UVB goes up, this should happen at proportionally higher densities and lower temperatures.  Thus the relatively narrow bands of number density where significant differences occur can be understood in this way.  

\begin{figure*}
    \centering
    \includegraphics[scale=0.4]{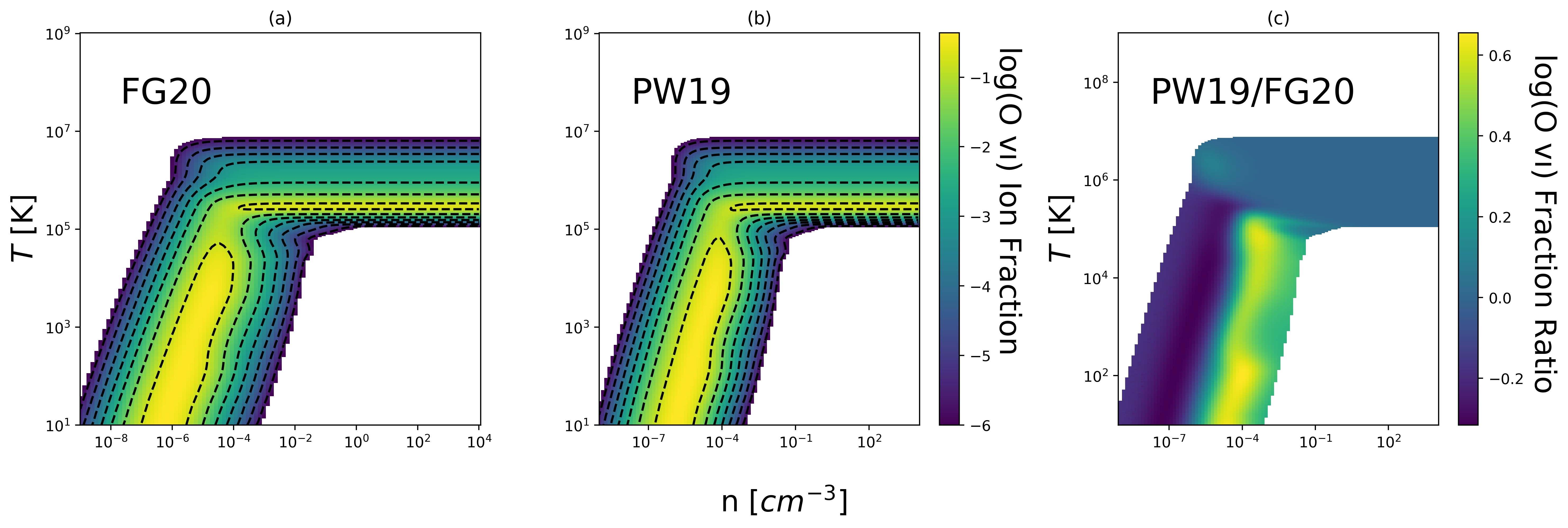}
    \caption{2D histogram showing the distribution of \ovi\ ion fraction over a range of gas density (x-axis) and temperature (y-axis). Panel (a) and panel (b) show the log-scale fraction of \ovi\ for FG20 and PW19 respectively. Panel (c) is the log ratio (PW19/FG20) of the two UVB ion fractions.}
    \label{fig:ion_frac}
\end{figure*}

Consider, for example, the differences observed in \ovi\ ion fractions between the FG20 and PW19 UVBs in Figure~\ref{fig:ion_frac}. This figure is a series of 2D histograms showing the ion fraction of \ovi\ for PW19 and FG20 at different gas densities and temperatures. The first two panels (starting from the left), panel (a) and panel (b), show the individual \ovi\ ion fractions for FG20 and PW19 respectively, while panel (c) shows the ratio between the two (PW19/FG20).

At lower number densities ($\log{n} < 5.5$) and temperatures ($\log{T} < 10^5 K$), these panels  show that the largest concentration of \ovi\ is centered around a diagonal line of gas density to temperature. At roughly $\log{n} = -5.5$ and $\log{T} = 10^5 K$ there is a very dramatic change in the ion fraction pattern in which the ion fraction becomes invariant with respect to number density. This represents the density where collisional ionization begins to dominate over photoionization. 

Now looking at the classification of absorber pairs (see Section~\ref{ssec:abs_cat}), we find that the vast majority of absorber pairs that were classified as a ``match'' tend to agree well between the two models being compared in terms of column density, gas density, and temperature. The most egregious outliers that we find do not fall into the ``match'' category. These outlying absorbers occupy different numbers of cells and positions along the LOS, and thus tend to be in the ``overlap'' category -- i.e., they are absorbers identified by the SALSA code that have some partial overlap in their physical extent, but only a small fraction of the total column density comes from the same region of gas. It is possible that a different approach to identifying absorbers would mitigate or remove these poorly-matched features entirely.

\subsection{Comparison to other work}

The vast majority of work examining the uncertainties in circumgalactic medium absorption line features come from observationally-focused efforts and thus focus on reverse modeling rather than the forward modeling approach that we have taken. One  study is from \citet{gibson2022}, who allow
the power-law slope of the metagalactic UV background to vary as a free parameter. In this study, the authors analyzed 34 observed absorbers with several different ions (\cii, \ciii, \mgii, \oi, \oii, \oiii, \oiv, \suiii, \siii, and \siiii). Doing this resulted in average uncertainties in CGM metal ion column densities to increase from 0.08 to 1.14 dex. 
Our work generally agrees with Gibson et al. in terms of the magnitude of uncertainty, with the majority of uncertainties in the species we consider (outside of \hi) remaining within 1 dex.
\citet{werk2014}, an observational paper from the COS survey, also experiments with slopes and intensities of ionization fields, choosing to vary the UVB between HM12 and a previous iteration of the HM12 model from 2001 that was not included in this study. They find that this results in an uncertainty of 0.3 dex in all derived gas parameters. This is significantly lower than the error we see in our work. However, this makes sense as these are comparisons between models of the same author which have less uncertainty than between models of different authors as seen in our work.

\cite{mallik2023} explores a similar scope of questions as those that we address in this work. These authors utilize both a forward and backward modeling approach and look at the impact of the metagalactic UV background on simulated absorption features and their subsequently inferred column densities rather than directly looking at the absorbers themselves. Although this leads to more sources of uncertainty it also produces results that are  more directly comparable to observation. Our work complements the findings of \cite{mallik2023} by showing that the uncertainty in metal absorbers exists not only through the forward and backward modeling approach taken in their paper, but by examining the impact of the UVB on the physical state of the gas.  This changes the ionization state of the gas, impacting the physical extent and overall quantity of each species.

\cite{acharya2022} finds similar results in their study. This work also employs reverse modeling techniques via the creation of ``toy'' absorbers, using CLOUDY then Bayesian MCMC and least-square differences to recover the \hi\ number density and metallicity.  They find similar results to our work, with the main difference being the authors had a much more significant focus on inferred physical quantities such as density and metalicity as opposed to the impact on column density. Their metalicity calculations originate from using several different ions, even varying their number and types of ions being used in their model. While this does introduce some difficulty in drawing direct comparisons between our works, there are a few notable instances in which there are some similarities between the works. Notably, their comparison between $N_{HI}$ between FG20 and PW19, they find that PW19 has a notably higher $N_{HI}$ than FG20, which agrees very well with results of this work.


Another study that analyzes the CGM at $z\approx2$, \citet{lehner2022}, analyzes the metallicity of CGM absorbers observed by KODIAQ-Z as compared to FOGGIE simulation data. To extract absorbers from the FOGGIE galaxy the authors also use the Trident and SALSA codes, using HM12 as the UV background for their analysis. Given the insights of our work the metallicity in their analysis would likely remain relatively unchanged, regardless of which UVB model was used. However, there would likely be some changes in the \hi\ column densities in the range of $\sim 10^{18}$--$10^{20}$ cm$^{-2}$, as the more recent UVB models predict significantly lower gas densities of \hi\ ($\approx$ 0.5 dex) in this column density range.

\subsection{Limitations and future work}

One of the primary limitations
of this work stems from our assumption that absorbers are physically contiguous systems, 
which is inherent to the absorber-finding algorithm embodied by the SALSA code. While this allows us to 
extract absorber column densities directly from the FOGGIE simulation data rather than estimating it by analyzing synthetic absorption spectra, it does not allow physically distinct structures with different line-of-sight velocities to contribute to a single absorption feature.  This means that there is a gap in realism between our absorbers and observed features.  That said, the goal of our work is not to be as physically realistic as possible -- our goal is
ultimately to compare the impact of different UVB models on a variety of commonly-observed ions in the circumgalactic medium, and the method we have chosen allows us to directly probe the physical impact of these backgrounds on the absorbing gas itself without any confusion relating to the blending of different physical features.

An additional limitation relates to the cutoff fraction and minimum column density variables used in the SALSA code's iterative absorber-finding algorithm  (see Section~\ref{sec:absorbers}).  While the impact of these parameter values were investigated and values were chosen that provided sensible results, there is still some room for ambiguity. To account for this, we used a large number of rays to account for any "missing" absorbers these parameters would eliminate, but they still will have some influence on the number of detected absorbers and final range of column densities.

Our method for matching absorbers between UVB inferences resulted in a decent amount of absorbers unsuitable for analysis. This was particularly detrimental to our analysis of \nv\. In particular, the use of somewhat arbitrary cell-based margin of error in the absorber categorization script (Section \ref{ssec:abs_cat}) is complicated by the varied physical sizes of each cell along and between the randomly-oriented ray objects (see Section~\ref{sec:extracting los}). 
Future versions of this matching algorithm should attempt to remedy these issues by instead setting an error region based on physical size or some other physical parameter rather than simulation cell counts. 

Finally, we note that our post-processing of the FOGGIE simulation data with different metagalactic ultraviolet backgrounds is not entirely physically consistent.
These simulations were run using the \citet{haart2012} UVB, and thus have values of density and temperature that evolve under that model assumption.  When we post-process the simulation data we use the baryon temperature, density, and metallicity fraction values that come from the dataset but we then calculate species ionization fractions using one of the four UV backgrounds.  This means that the gas values are not completely consistent with the UV background used in post-processing.  A more nominally self-consistent way to approach this would be to run multiple simulations, each with different UV backgrounds.  This would have some impact on the dynamical evolution of the system, however, and due to the inherently chaotic nature of galaxy evolution this would mean that we would not be able to find identical physical structures in each simulation.  As a result, the approach that we have taken is the only realistic way to perform absorber-by-absorber matching.

There are several additional avenues of future work.  One concerns the variation of elemental abundances in the circumgalactic medium.  In this work we have implicitly assumed that all gas in the circumgalactic medium, regardless of metallicity, has a Solar abundance pattern.  This is not always true, particularly at lower metallicities and at high redshifts.  In fact, we would expect that the (non-hydrogen) elements that are commonly observed in the CGM would have some variation in their relative abundance -- carbon comes from both exploding massive stars (i.e., Type II supernovae) and from low-mass stars, whereas oxygen and magnesium come entirely from massive stars.  Even the elements that come from the same sources will not necessary end up with precisely the same ratios. Furthermore, there is a delay between enrichment from Type II and Type Ia supernovae that will significantly impact the relative contributions of various alpha and iron peak elements. This uncertainty in the chemical composition of the circumgalactic medium will itself contribute to systematic uncertainties in the determination of its overall metallicity beyond those relating to the uncertainty in UV background.
Future work will revisit the experiment presented in this paper using different abundance patterns that could potentially more accurately represent that which might be observed in the CGM.

We also note that we do not consider the inclusion of the variation of the UV background near galaxies due to the massive stars in that galaxy. This is potentially important in calculating the ionization state of gas that is in the inner CGM \citep[see, e.g.,][]{werk2016}, and it likely also depend on the azimuthal angle---i.e., UV light from stars is going to preferentially escape along galactic poles rather than along the equatorial plane. This is also a clear area for future work. 

Finally, including the line-of-sight velocity of absorbers in our absorber-finding algorithm is an additional step towards direct comparison to observational results, as this is one of the major limitations of the SALSA algorithm's approach. 

\section{Conclusions} 
\label{sec:conclusion}

In this paper we examine the uncertainties in the column densities of physically-contiguous absorption systems that come from variation in assumptions relating to the metagalactic ultraviolet background.  We do this using a ``forward modeling'' approach where ``pencil beams'' of physical quantities are extracted from the circumgalactic medium of high resolution cosmological simulations of galaxy evolution from the FOGGIE project, different UV background models are applied to calculate the densities of commonly-observed ions (\hi, \siii, \siiii, \siiv, \civ, \nv, and \ovi), and physically contiguous structures are identified.  We perform comparisons between pairs of ultraviolet backgrounds, categorize the outcomes, and examine the relationship between the degree of agreement between absorbers of different column densities as a function of UV background and the thermodynamic state of the underlying gas.

From our analysis, we found several key results:

\begin{enumerate} 
    \item There exist significant absorber column density differences based on the choice of ultraviolet background, with notable trends between model generations, model families, and across ions of different ionization energies. These differences can range from \~0.5 dex to as much as \~ 6 dex in some instances.
    
    \item With the notable exception of \hi, differences between model's absorber densities tend to increase with increasing ionization energy. This is characterized in both the total column density comparison (Fig. \ref{fig:tot_col_dens}) by the increasing density offsets between models with increasing ionization energy, in the direct UVB comparison (Fig. \ref{fig:1:1comparison}, \ref{fig:fg_v_fg}, \ref{fig:hm_v_p}, \ref{fig:fg_vs_p}), and the larger scatter of absorber column density differences with higher ionization energy ions.
    
    \item Many of the observed trends in the absorber comparisons can be understood through examining differences in the UVB spectra (i.e., regions with significant differences between UVBs tend to have differences in the column densities of ions with ionization energies in that energy region).
    
    \item In general, species where the magnitude of the UVB flux at the ionization potential vary significantly (such as \hi, \nv, or \ovi) show significant offsets between the total amount of each ion, whereas species where the UVB flux magnitudes are much closer at the ionization energy (e.g., \siiii) tend to show correspondingly smaller systematic offsets
    
    \item Absorbers with matching spatial positions tend to have more similar column densities. Only small ($\lesssim 0.1$ dex) differences between matched absorbers are seen when comparing within the Faucher-Gigu\`ere family of models. When comparing the models within the Haardt-Madau-Puchwein family as well as between the two families of models, the most striking differences are seen in absorbers of relatively low column densities (particularly for \hi, \civ, and \ovi) where variations in the UV background would be most impactful.
    
    \item The PW19 model produces significantly more \hi, \ciii, \civ, \nv, and \ovi\ compared to both its older HM12 parent model and to FG20, it has absorbers with significantly higher densities of \hi, \civ, \nv, and \ovi. 
    
\end{enumerate}

Having utilized a forward-modeling approach for generating synthetic observations of the CGM, our work serves as a compliment to previous work, handling the issue of making comparisons between synthetic and real observations in a novel way. As this approach continues to grow in range and accuracy, the results of this work, along with backward-modeling approaches, can be used to vastly improve the effectiveness of CGM simulation.

\begin{acknowledgments}
The authors thank E.~Puchwein for sharing UV background data with us.  BWO thanks Rongmon Bordoloi, Christopher Howk, Nicolas Lehner, and Jess Werk for useful discussions.
ET, CK, and BWO acknowledge support from NSF grants \#1908109 and \#2106575 and NASA ATP grants NNX15AP39G and 80NSSC18K1105. 
This work used resources provided by the Advanced Cyberinfrastructure Coordination Ecosystem: Services \& Support (ACCESS), which is supported by National Science Foundation grant number TG-AST090040, as well as the resources of the Michigan State University High Performance Computing Center (operated by the MSU Institute for Cyber-Enabled Research).   Computations described in this work were performed using the publicly-available yt \citep{yt_2011ApJS..192....9T},   Trident \citep{trident-2017ApJ...847...59H}, and SALSA~\citep{Salsa2020JOSS....5.2581B} codes, which are the products of the collaborative effort of many independent  scientists  from  numerous  institutions  around  the world. This paper uses data from the Figuring Out Gas and Galaxies In Enzo project \citep[FOGGIE; ][and \url{https://foggie.science}]{peeples2019}.  

\end{acknowledgments}

\software{astropy \citep{2013A&A...558A..33A,2018AJ....156..123A},  
yt \citep{yt_2011ApJS..192....9T}, Trident~\citep{trident-2017ApJ...847...59H}, SALSA~\citep{Salsa2020JOSS....5.2581B}, CLOUDY~\citep{2013RMxAA..49..137F}
          }

\bibliography{sample631}{}
\bibliographystyle{aasjournal}

\end{document}